\documentclass[11pt]{article}
\usepackage{amsfonts}
\usepackage{listings}
\usepackage{amssymb}
\usepackage{amsmath}
\usepackage{amsxtra}
\usepackage{graphicx}
\usepackage{psfrag}
\usepackage{mathrsfs}
\usepackage{natbib}
\usepackage{stmaryrd}
\usepackage{subfigure}
\usepackage{tikz}
\usepackage{empheq}
\usepackage[normalem]{ulem}

\begin{document}
\def\BGamma{\mbox{\boldmath$\Gamma$}}
\def\BDelta{\mbox{\boldmath$\Delta$}}
\def\BTheta{\mbox{\boldmath$\Theta$}}
\def\BLambda{\mbox{\boldmath$\Lambda$}}
\def\BXi{\mbox{\boldmath$\Xi$}}
\def\BPi{\mbox{\boldmath$\Pi$}}
\def\BSigma{\mbox{\boldmath$\Sigma$}}
\def\BUpsilon{\mbox{\boldmath$\Upsilon$}}
\def\BPhi{\mbox{\boldmath$\Phi$}}
\def\BPsi{\mbox{\boldmath$\Psi$}}
\def\BOmega{\mbox{\boldmath$\Omega$}}
\def\Balpha{\mbox{\boldmath$\alpha$}}
\def\Bbeta{\mbox{\boldmath$\beta$}}
\def\Bgamma{\mbox{\boldmath$\gamma$}}
\def\Bdelta{\mbox{\boldmath$\delta$}}
\def\Bepsilon{\mbox{\boldmath$\epsilon$}}
\def\Bzeta{\mbox{\boldmath$\zeta$}}
\def\Beta{\mbox{\boldmath$\eta$}}
\def\Btheta{\mbox{\boldmath$\theta$}}
\def\Biota{\mbox{\boldmath$\iota$}}
\def\Bkappa{\mbox{\boldmath$\kappa$}}
\def\Blambda{\mbox{\boldmath$\lambda$}}
\def\Bmu{\mbox{\boldmath$\mu$}}
\def\Bnu{\mbox{\boldmath$\nu$}}
\def\Bxi{\mbox{\boldmath$\xi$}}
\def\Bpi{\mbox{\boldmath$\pi$}}
\def\Brho{\mbox{\boldmath$\rho$}}
\def\Bsigma{\mbox{\boldmath$\sigma$}}
\def\Btau{\mbox{\boldmath$\tau$}}
\def\Bupsilon{\mbox{\boldmath$\upsilon$}}
\def\Bphi{\mbox{\boldmath$\phi$}}
\def\Bchi{\mbox{\boldmath$\chi$}}
\def\Bpsi{\mbox{\boldmath$\psi$}}
\def\Bomega{\mbox{\boldmath$\omega$}}
\def\Bvarepsilon{\mbox{\boldmath$\varepsilon$}}
\def\Bvartheta{\mbox{\boldmath$\vartheta$}}
\def\Bvarpi{\mbox{\boldmath$\varpi$}}
\def\Bvarrho{\mbox{\boldmath$\varrho$}}
\def\Bvarsigma{\mbox{\boldmath$\varsigma$}}
\def\Bvarphi{\mbox{\boldmath$\varphi$}}
\def\bone{\mbox{\boldmath$1$}}
\def\bzero{\mbox{\boldmath$0$}}
\def\bnabla{\mbox{\boldmath$\nabla$}}
\def\bvarepsilon{\mbox{\boldmath$\varepsilon$}}
\def\bA{\mbox{\boldmath$ A$}}
\def\bB{\mbox{\boldmath$ B$}}
\def\bC{\mbox{\boldmath$ C$}}
\def\bD{\mbox{\boldmath$ D$}}
\def\bE{\mbox{\boldmath$ E$}}
\def\bF{\mbox{\boldmath$ F$}}
\def\bG{\mbox{\boldmath$ G$}}
\def\bH{\mbox{\boldmath$ H$}}
\def\bI{\mbox{\boldmath$ I$}}
\def\bJ{\mbox{\boldmath$ J$}}
\def\bK{\mbox{\boldmath$ K$}}
\def\bL{\mbox{\boldmath$ L$}}
\def\bM{\mbox{\boldmath$ M$}}
\def\bN{\mbox{\boldmath$ N$}}
\def\bO{\mbox{\boldmath$ O$}}
\def\bP{\mbox{\boldmath$ P$}}
\def\bQ{\mbox{\boldmath$ Q$}}
\def\bR{\mbox{\boldmath$ R$}}
\def\bS{\mbox{\boldmath$ S$}}
\def\bT{\mbox{\boldmath$ T$}}
\def\bU{\mbox{\boldmath$ U$}}
\def\bV{\mbox{\boldmath$ V$}}
\def\bW{\mbox{\boldmath$ W$}}
\def\bX{\mbox{\boldmath$ X$}}
\def\bY{\mbox{\boldmath$ Y$}}
\def\bZ{\mbox{\boldmath$ Z$}}
\def\ba{\mbox{\boldmath$ a$}}
\def\bb{\mbox{\boldmath$ b$}}
\def\bc{\mbox{\boldmath$ c$}}
\def\bd{\mbox{\boldmath$ d$}}
\def\be{\mbox{\boldmath$ e$}}
\def\bff{\mbox{\boldmath$ f$}}
\def\bg{\mbox{\boldmath$ g$}}
\def\bh{\mbox{\boldmath$ h$}}
\def\bi{\mbox{\boldmath$ i$}}
\def\bj{\mbox{\boldmath$ j$}}
\def\bk{\mbox{\boldmath$ k$}}
\def\bl{\mbox{\boldmath$ l$}}
\def\bm{\mbox{\boldmath$ m$}}
\def\bn{\mbox{\boldmath$ n$}}
\def\bo{\mbox{\boldmath$ o$}}
\def\bp{\mbox{\boldmath$ p$}}
\def\bq{\mbox{\boldmath$ q$}}
\def\br{\mbox{\boldmath$ r$}}
\def\bs{\mbox{\boldmath$ s$}}
\def\bt{\mbox{\boldmath$ t$}}
\def\bu{\mbox{\boldmath$ u$}}
\def\bv{\mbox{\boldmath$ v$}}
\def\bw{\mbox{\boldmath$ w$}}
\def\bx{\mbox{\boldmath$ x$}}
\def\by{\mbox{\boldmath$ y$}}
\def\bz{\mbox{\boldmath$ z$}}
\newcommand*\mycirc[1]{%
  \begin{tikzpicture}
    \node[draw,circle,inner sep=1pt] {#1};
  \end{tikzpicture}
}

\title{Parallel Level set algorithm with MPI and accelerated on GPU }
\author{Zhenlin Wang \\ University of Michigan, Ann Arbor}
\maketitle
\abstract{Level set method has been used to capture interface motion. Narrow band algorithm is applied to localize the solving of level-set PDE on global domain to a tube around interface. Due to the unknown evolving interface, narrow band algorithm brings load balance problem for parallelizing computing. This work presents a tool for evenly distributing work loads on CPU cores. On the other hand, numerically solving level-set PDE only needs simple operations but on large grid points. This work also presents a GPU acceleration for solving level-set PDE using finite difference method. }
%
%
\section{Introduction of level set method}
Level set method has been used to capture interface motion. The central ideas of level set methods are first representing the interface as the zero level set of a higher dimensional function, second embedding the interface's velocity $F$ to this higher dimensional level set function. The details of level set method could be found at (\cite{Sethian2000}). The essential formulations are discussed and summarized here. Mathematically to track a moving closed hyper-surface $\Gamma(t)$, let $\pm d$ be the signed distance to the interface, We can define 
\begin{align}
\phi(\bx,t=0)=\pm d
\label{eq:signedDistance}
\end{align}
then an differential equation can be obtained for the evolution of $\phi$,
\begin{align}
\phi_t+F|\nabla \phi|=0
\end{align}
with initial value shown by equation (\ref{eq:signedDistance}).
Then the position of the evolving front $\Gamma$ at time $t$ is given by the zero level set $\phi(\bx,t)=0$.

For solving this initial value problem numerically, upwind viscosity schemes \citep{finiteDPatty2010} are stable and used in this work. Specifically a first order scheme is used
\begin{align}
\phi^{n+1}_{ijk}=\phi^{n}_{ijk}-\Delta t[\max(F_{ijk},0)\nabla^{+} + \min(F_{ijk},0)\nabla^-]
\end{align}
where
\begin{align}
\nabla^+ &= \left[ 
\begin{array}{l}
\max(D^{-x}_{ijk},0)^2+\min(D^{+x}_{ijk},0)^2+\\
\max(D^{-y}_{ijk},0)^2+\min(D^{+y}_{ijk},0)^2+\\
\max(D^{-z}_{ijk},0)^2+\min(D^{+z}_{ijk},0)^2+
\end{array}\right]^{1/2}
\end{align}

\begin{align}
\nabla^- &= \left[ 
\begin{array}{l}
\max(D^{+x}_{ijk},0)^2+\min(D^{-x}_{ijk},0)^2+\\
\max(D^{+y}_{ijk},0)^2+\min(D^{-y}_{ijk},0)^2+\\
\max(D^{+z}_{ijk},0)^2+\min(D^{-z}_{ijk},0)^2+
\end{array}\right]^{1/2}
\end{align}
\subsection{local level set: Narrow band algorithm}
The problem could be solved globally over whole domain, while it is quite expensive with computational labor ($O(n^3)$) per time step. However we are only interesting in zero level set $\phi(\bx,t)=0$. Narrow band algorithm is introduced to localize the level set problem to a narrow tube around the interface $\Gamma$, the zero level set. The details could be found at \citep{localLevelSetZhao1999}.The algorithm is summarized below.

\begin{figure}[hbtp]
\centering
\includegraphics[scale=0.6]{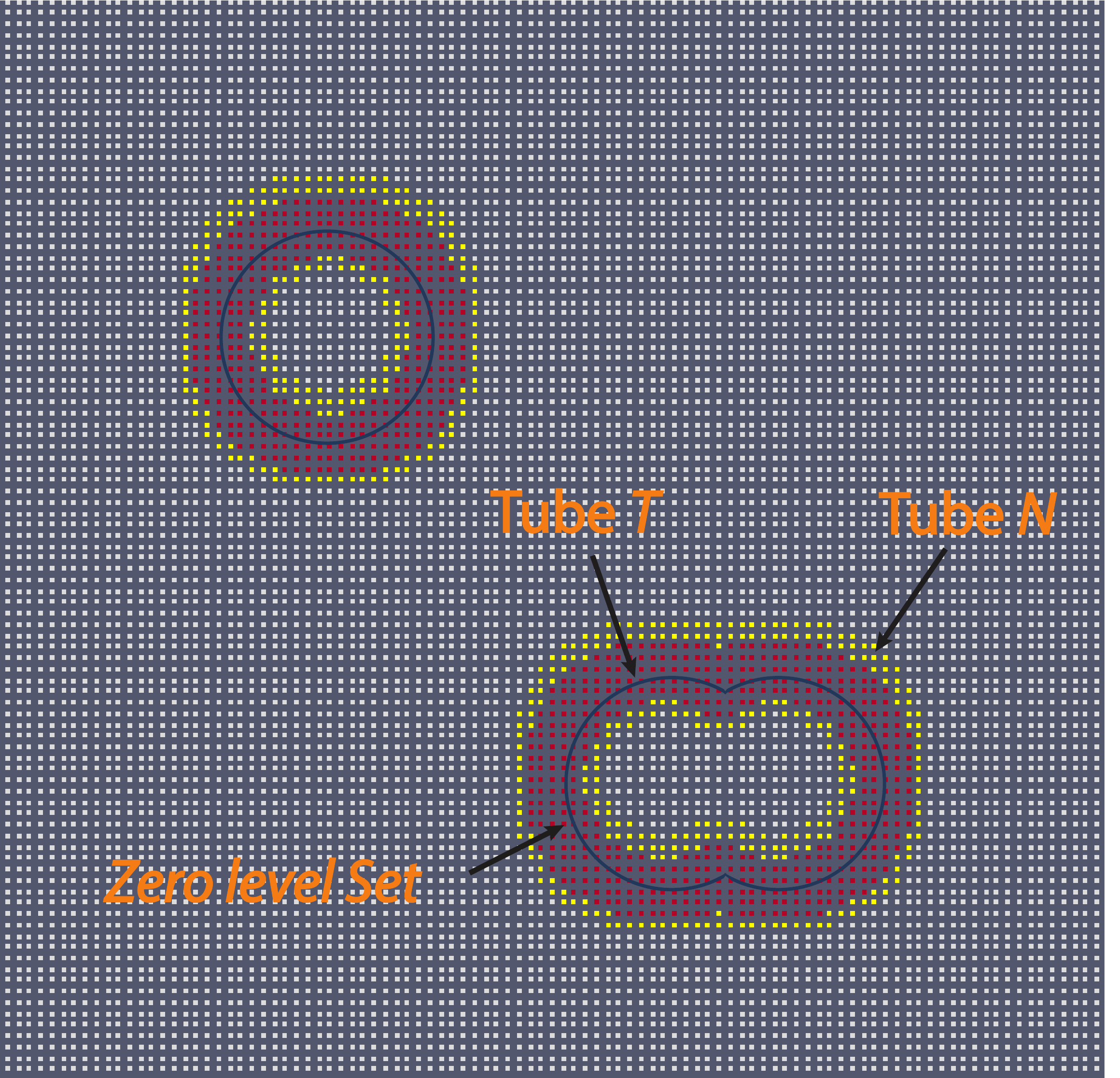}
\caption{Red points are in tube $T$, red and yellow points are in tube $N$. $\phi$ is updated on tube $T$ and reinitialized on tube $N$. }
\label{fig:narrowBand}
\end{figure}

The algorithm is easy to demonstrated by figure (\ref{fig:narrowBand}). Let $\gamma>\beta>0$ be two constants which are comparable to grid size $\Delta x$. Initially for a given front $\Gamma^0$, we define a tube with width $\gamma$ by
\begin{align}
T^0=\{x:|\phi^0(x)|<\gamma\}
\label{eq:T}
\end{align}
Let $c$ be a cut-off function
\begin{align}
c(\phi) &= \left\{
\begin{array}{ll}
1& |\phi|\le \beta \\
(|\phi|-\gamma)^2(2|\phi|+\gamma-3\beta)/(\gamma-\beta)^3 & \beta<|\phi|\le \gamma \\
0 & |\phi|\ge \gamma
\end{array}\right.
\end{align}
We update $\gamma^0$ by solving the following equation
\begin{align}
\phi_t+c(\phi)F|\nabla \phi|=0
\label{eq:narrowBand}
\end{align}
on $T^0$ with initial value given by equation (\ref{eq:signedDistance}). The new location of front is given by $\Gamma^1=\{x:\bar{\phi}^1(x)=0\}$. Let $d^1$ which is not known currently be the signed distance to $\Gamma^1$. To move the front we need the shifted tube
\begin{align}
T^1=\{x:|d^1(x)|<\gamma\}
\end{align}
and construct a new level set function such that
\begin{align}
\phi^1(x) &= \left\{
\begin{array}{ll}
-\gamma & d^1(x)<-\gamma \\
d^1(x) & |d^1(x)|\le\gamma \\
\gamma & d^1(x)>\gamma
\end{array}\right.
\label{eq:updatePhi}
\end{align}
Then the question is how to compute $d^1(x)$, in other word how to reinitialize signed distance function. The reinitialization must be performed on a region that contains $T^1$, since the front moves less than one grid point, we can choose this this region to be
\begin{align}
N^0=\{x:|\phi^0(x\pm \Delta x)|<\gamma\}
\label{eq:N}
\end{align}
On this region, the following Hamilton-Jacobi type equation,
\begin{align}
d_\tau+S(d)(|\nabla d|-1)=0 \label{eq:reIniti}\\
d(x,0)=\bar{\phi}(x) \nonumber
\end{align}
is solved to steady state. For stability and smoothing, $S(d)$ is choosed as
\begin{align}
S=\frac{d}{\sqrt{d^2+|Dd|^2\Delta x^2}}
\end{align}
And same upwind viscosity schemes are applied for approximating the gradient term.

The algorithm for narrow band method is summarized as \\
-----------------------------------------\\
\textbf{Step 0}: Initialization. Give initial interface $\Gamma^0$, initialize $\phi^0$ by signed distance function to the interface.\\
\textbf{Step 1}: Narrow band. Set tubes $T$ and $N$ by equations (\ref{eq:T}) and (\ref{eq:N}).\\
\textbf{Step 2}: Advance. Update $\phi$ in tube $T$ for one time step to get $\bar{\phi}$ by equation (\ref{eq:narrowBand}).\\
\textbf{Step 3}: Reinitialization. Apply the reinitialization step to $\bar{\phi}$ on the tube $N$ by equation (\ref{eq:reIniti}), and define new $\phi$ by equation (\ref{eq:updatePhi}).\\
\textbf{Step 4}: Output results. Go back to step 1.\\
-----------------------------------------\\
\section{parallelization on distributed memory}
Consider the narrow band algorithm, constructing the narrow band needs iterating over all grid points, while updating and reinitializing $\phi$ is on the narrow tube. The technique of decomposing the problem is illustrated below by a 2D grid, and could be easily extended for 3D grid. 

\begin{figure}
\centering
\includegraphics[scale=0.5]{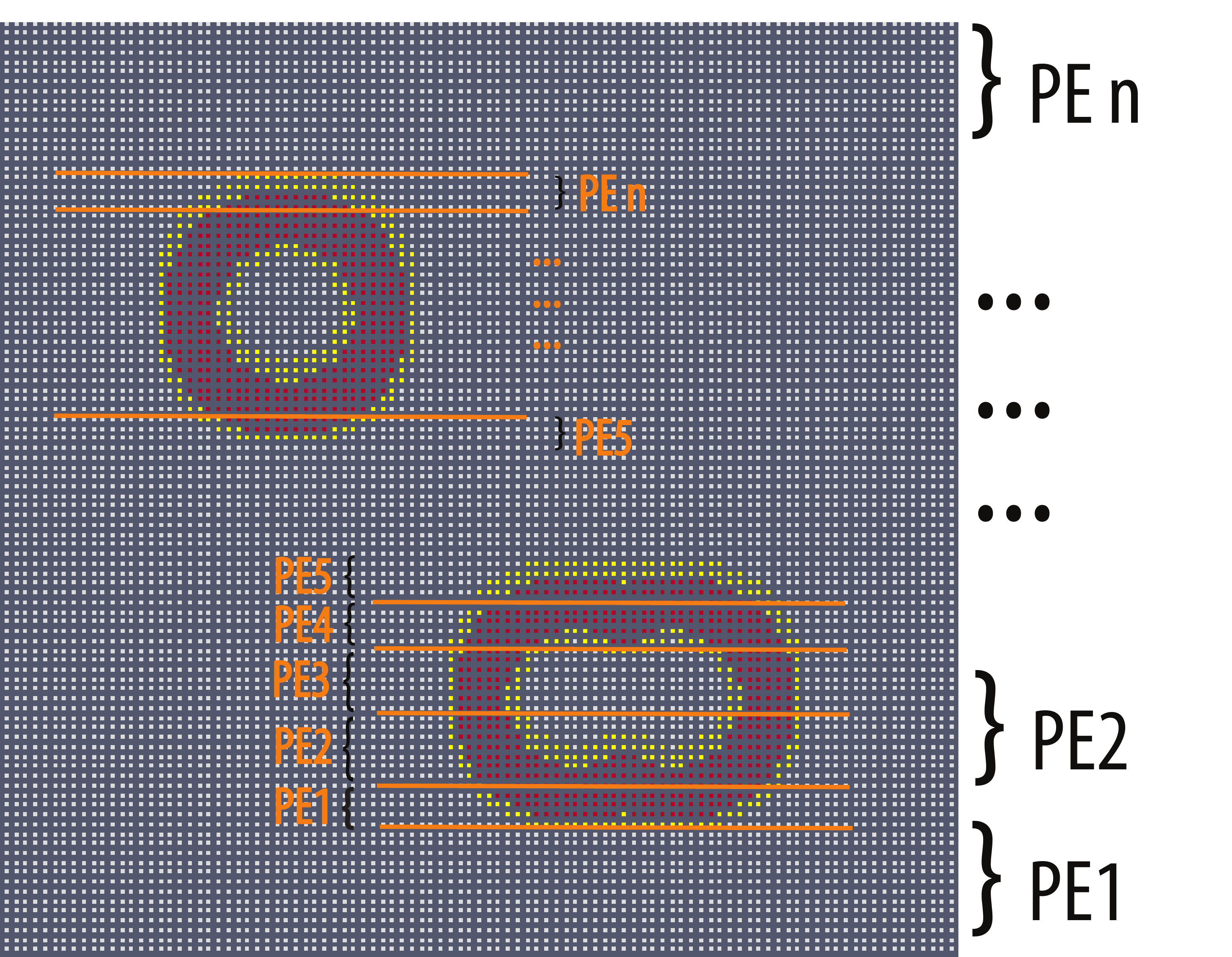}
\caption{The whole grid is decomposed globally by row. The tube is dynamically decomposed each time after tube is update.}
\label{fig:partition}
\end{figure}

Let's consider a grid with $x_1*x_2$ points and we have n cores. For operation over whole domain (e.g. constructing the narrow band) it is easy to decompose the grid is by row on 2D or by one axis for 3D shown in figure (\ref {fig:partition}). Each cores will contain $x_2/p$ rows with two extra ghost rows (the first and last cores only need one ghost row). This global decomposition is for constructing the narrow band which needs operating all grid points. 

After the narrow tube has beed constructed, all points in tube (active points) are stored in a array. We want to evenly distribute the active points to different cores. And before reinitialization to $\bar{\phi}$ on the tube $N$, cores each to communicate to update the $\phi$ of ghost cell. It is very hard to track the shape of the tube and decompose the tube evenly based on its specific shape. To overcome the problem, the tube is also decomposed by row on 2D or by one axis for 3D, but the number of rows assigned to one core is determined as following. Suppose there are $M$ points in the tube and $n$ cores. We want each core has active points close to $M/n$, and also assigning active points row by row. Suppose there are $S$ points currently, and next row contains $c$ active points. If
\begin{align}
|S-\frac{M}{n}|>|S+c-\frac{M}{n}| \Rightarrow S<\frac{M}{n}-\frac{c}{2}
\end{align}
that is, adding active points of next row will make total points closer to $M/n$, next row is included for current core, otherwise begin to assign rows for next core.  Using this allocation routine, firstly active points are allocated by row which means each core only need to communicate with its adjacent core to update ghost cell before reinitialization step, and secondly each core has active points closest to $M/n$, thirdly it is very easy to handle different shape of tube and multiple tubes, and does not need to track the interface evolving.  
In the code, result of $\phi$ at each time step will be outputted as vts file. Writing result itself cannot be parallelized, but results for several time step could be outputted simultaneously. One strategy is all cores do calculation for one time step, then one core do outputting and continue for next time step calculation. Obviously this will make all cores wait for the one doing outputting and slow down the simulation. Another way is let all cores keep doing calculation and store the results, then all cores doing outputting finally. This is fastest but will cause memory accumulation. In this work, the strategy is let one cores alway do outputting while all other cores keep doing calculation. After all calculations are finished, all processors start outputting remaining results.

To minimize communication between two cores, all data is packed into one array and send it once. 

\subsection{Manager-workers paradigm}
For implementation, "manager-workers" paradigm is used. The parallelized narrow band algorithm is shown following:\\
-----------------------------------------\\
\textbf{Step 1}: Narrow band. Manager processor send information of $\phi$ with ghost cell to workers by global decomposition, workers return active points to manager processor which constructs the narrow band afterwards.\\
\textbf{Step 2}: Advance.  Manager processor dynamically allocates active points to workers, workers Update $\phi$ in partial tube $T$ for one time step to get $\bar{\phi}$ by equation (\ref{eq:narrowBand}).\\
\textbf{Step 3}: Reinitialization. Workers communicate with its adjacent processor to update ghost cell and apply the reinitialization step to $\bar{\phi}$ on the partial tube $N$ by equation (\ref{eq:reIniti}), define new $\phi$ by equation (\ref{eq:updatePhi}). Return the new $\phi$ to manager and wait.\\
\textbf{Step 4}: Output results. Manager processor using nonblocking "MPI\_Iprobe" to checks the status of processor for outputting result. If it is free manager processor send all results to it and clear the results queue. If it is still busy on outputting,  manager processor push results of this time step to queue. Go back to step 1.\\
\textbf{processor for outputting:} After finish writing all results it received, send "finish writing" message and wait.\\
\textbf{termination:} After all calculation finished, manager processor evenly distributes all results in queue to all free processors, after receives all "finish writing" message, sends "stop" message to break waiting of workers. \\
 -----------------------------------------\\
 The types of working (e.g. outputting data, doing finite difference calculation or setting narrow band) are conveyed by "TAG". Then workers will do specific work according to the types it received shown as following code. \\
  -----------------------------------------\\
\begin{lstlisting}
 while(true){
	int flag;
	int number_amount;
	MPI_Probe(0, MPI_ANY_TAG, MPI_COMM_WORLD, &status);
	if(status.MPI_TAG==stopWork) {
		break;
	}
	if(status.MPI_TAG==tag_output) {
		MPI_Get_count(&status, MPI_DOUBLE, &number_amount);
		output(number_amount);
	}
	if(status.MPI_TAG==finiteD){
		MPI_Get_count(&status, MPI_DOUBLE, &number_amount);
		finiteDifference(number_amount);
		reInitialization();
	}			
	if(status.MPI_TAG==work_setNarrowBand){
		MPI_Get_count(&status, MPI_DOUBLE, &number_amount);
		setNarrowBand(number_amount);
	}		
}
\end{lstlisting}
 -----------------------------------------\\
 By doing so, the code is easy to be extended for more complicated arrangement of worker processors doing different jobs.
\section{Acceleration by GPU}
numerically solving level-set PDE using finite difference method only needs simple operations but on large grid points. This part of work is suitable for GPU acceleration. On GPU, each grid point could be processed by one thread. This will greatly accelerate the calculation. To do so, one possible partition could be\\
 -----------------------------------------\\
\begin{lstlisting}
   dim3 threadsPerBlock(num_point[0],1,1);
   dim3 numBlcoks(num_point[1],num_point[2]);
\end{lstlisting}
 -----------------------------------------\\
then, the positions (x, y, z) of one grid point and its index of the data array could be easily accessed by\\
 -----------------------------------------\\
\begin{lstlisting}
   int x=threadIdx.x;
   int y=blockIdx.x;
   int z=blockIdx.y;
   int index=z*blockDim.x*gridDim.x+y*blockDim.x+x;
\end{lstlisting}
 -----------------------------------------\\
To minimize communication between CPU and GPU, we let GPU finish $m$ time steps calculation and then return all $m$ steps results to CPU. $m$ need to be tuned based on memory size of GPU. Since to update $\phi$, it needs to access $\phi$ at last time step, threads need to be synchronized before next time step calculation begins. However there is no global synchronization for all threads on CUDA ("\_\_syncthreads()" only synchronizes all threads within a block), multiple kernel invocations have to be applied. In this work, no CPU parallelization for outputting when GPU returns results to CPU.

\section{Numerical results}
I present two simulations with 10 time steps, one with grid points:101*101*101 and one with grid points:201*201*201 which is 8 times larger than previous one. Both simulations are carried by serial code, parallel code using 3, 4 ,8, 16 cores (Flux system), and on GPU (Bridge system). For each cases, I compared the run time with outputting results and without outputting results. The runtimes are summarized in table (\ref{tb:101}) and (\ref{tb:201}). 
\begin{table}[]
\centering
grid points:101*101*101\\ 
\begin{tabular}{c|c|c|c|c|c|c}
\hline
&serial code& 3 cores & 4 cores& 8 cores& 16 cores& GPU \\ \hline
without output /sec & 1.38e+02 &1.12e+02 & 5.60e+01 & 2.46e+01 & 1.91e+01 &  4.00e-02  \\ \hline
with output /sec&1.64e+02 &1.14e+02  &6.00e+01&2.73e+01 &2.08e+02 & 1.85e+01 
\end{tabular}
\caption{}
\label{tb:101}
\end{table}

\begin{figure}
\centering
\includegraphics[scale=0.6]{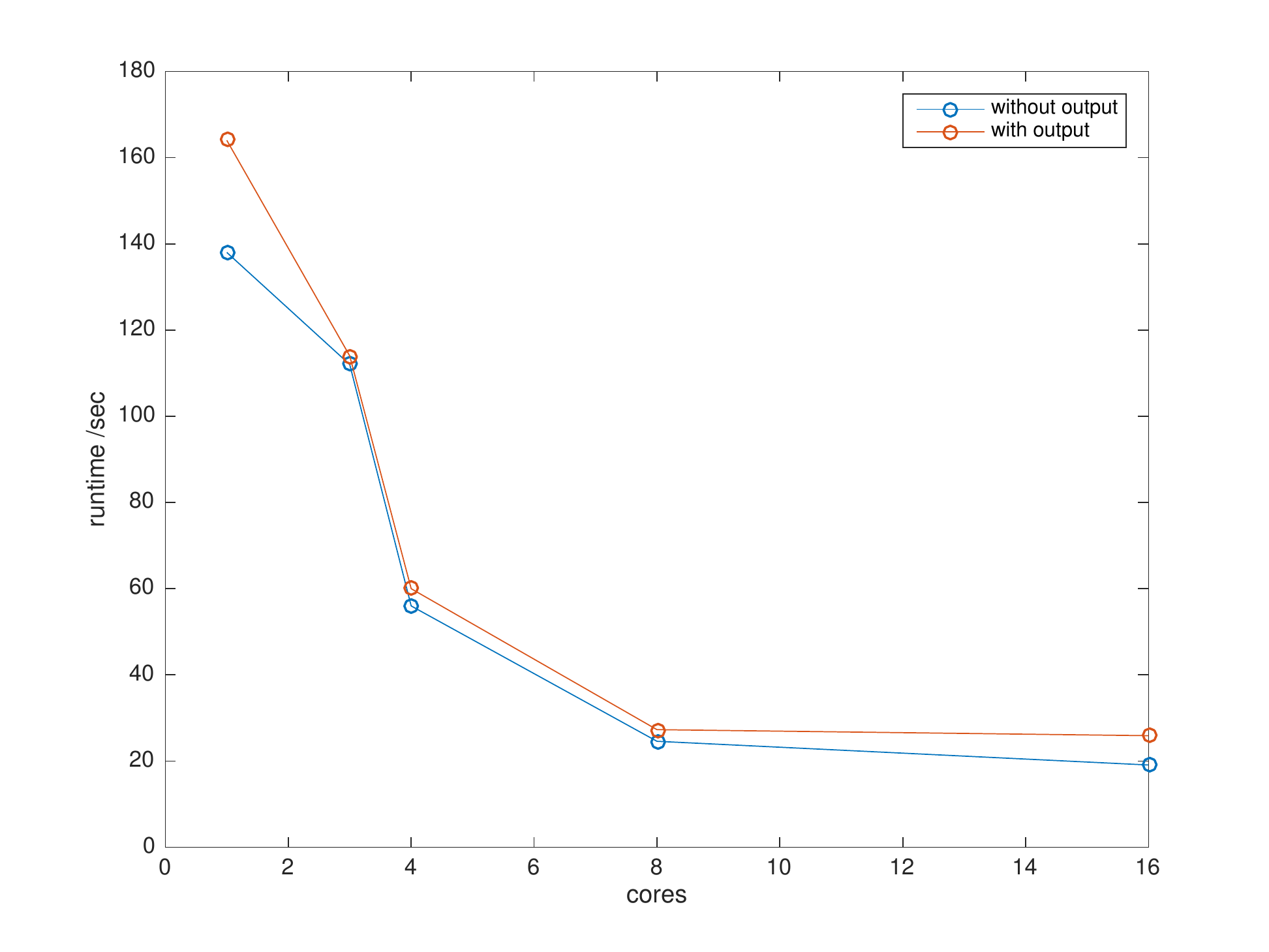}
\caption{runtime for grid points 101*101*101 on CPUs.}
\label{fig:runtime101}
\end{figure}

\begin{figure}
\centering
\includegraphics[scale=0.6]{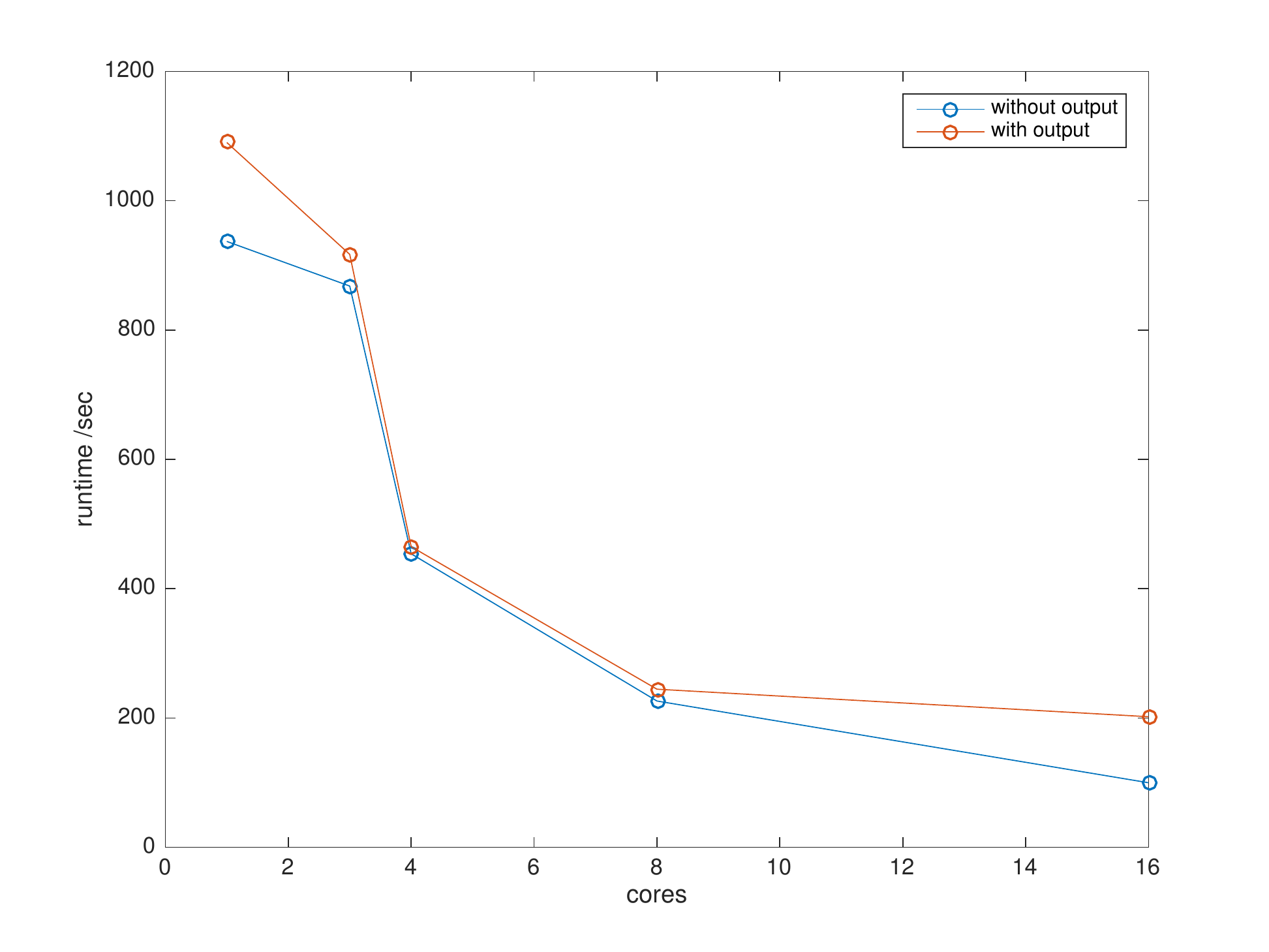}
\caption{runtime for grid points 201*201*201 on CPUs.}
\label{fig:runtime201}
\end{figure}

\begin{table}[]
\centering
grid points:201*201*201\\ 
\begin{tabular}{c|c|c|c|c|c|c}
\hline
&serial code& 3 cores & 4 cores& 8 cores& 16 cores& GPU \\ \hline
without output /sec &  9.37e+02 &8.68e+02 & 4.54e+02 & 2.26e+02 &  9.99e+01 &  3.60e-01  \\ \hline
with output /sec&1.09e+03 &9.17e+02  &4.65e+02& 2.45e+02  & 2.02e+02 &1.42e+02 
\end{tabular}
\caption{}
\label{tb:201}
\end{table}
From the serial code results, we found outputting results only take small portion ($\sim$16\%) of total time time. After parallelization the processor doing outputting always finished before other processors finish the calculation, so runtime without outputting results only slightly longer than that with outputting results. Problem with grid points 201*201*201 is 8 times larger than that with grid points 101*101*101, and the runtime for serial code and parallel code both properly scaled by problem size. For parallel code, when cores are increased, the runtime is scaled properly when cores smaller than 8 shown in figure (\ref{fig:runtime101}) and (\ref{fig:runtime201}). Note since Manager-workers paradigm is used one core will be excluded for manager. 

When GPU is used, calculation is speedup greatly. Compared the calculation time (without outputting), GPU is about 477 times faster than 16 cores for 101*101*101 grid points and about 300 times faster for 201*201*201 grid points. For runtime with outputting,  GPU also achieves fastest runtime even with serial outputting.

\section{Conclusion}
When solving level set problem on whole domain, it is computational expensive with labor ($O(n^3)$) per time step. But the problem is very easy to decompose which should achieve good scaling. Narrow band method will reduce the problem to  ($O(n^2)$) per time step, however it is harder to decompose the problem properly. And since computational complexity is reduced by one order, the overhead cost becomes significant and may causes bad scaling when use more cores. GPU is suitable for finite different calculation on large grid and greatly accelerates calculation. For faster simulation, GPU and CPU cores can work together, such as GPU doing  calculation and return several time steps results, then CPU cores output the results concurrently.  

The code is available at https://github.com/wzhenlin/parallel-levelSet.

\bibliographystyle{elsart-harv}
\bibliography{references}
\end{document}